\documentclass{PoS}

\newcommand{\gppr}{\stackrel{>}{\scriptstyle \sim}}
\newcommand{\lppr}{\stackrel{<}{\scriptstyle \sim}}

\title{UHE Cosmic Rays and AGN Jets}

\ShortTitle{UHE Cosmic Rays and AGN Jets}

\author{\speaker{Frank M. Rieger}\\
         ZAH, Institute of Theoretical Astrophysics, University of Heidelberg \\
         Philosophenweg 12, 69120 Heidelberg, Germany\\
        E-mail: \email{f.rieger@uni-heidelberg.de}}

\abstract{Active Galactic Nuclei (AGN) and their relativistic jets are believed to be potential sites of 
ultra-high-energy (UHE) cosmic ray acceleration. This paper reviews basic observational findings as 
well as requirements on source energetics, and then discusses the relevance of different acceleration 
sites and mechanisms, such as black hole gap, shock in back-flows or jet shear acceleration. When 
put in context, the result suggests that Fermi-type particle acceleration at trans-relativistic shocks 
and/or in shearing, relativistic flows offers the most promising framework for UHECR production in AGN.
Truly deciphering the astrophysical sources of UHECRs, however, still needs improved statistical 
information on arrival directions and source correlations.}

\FullConference{High Energy Phenomena in Relativistic Outflows VII - HEPRO VII\\
		9-12 July 2019\\
		Facultat de Fisica, Universitat de Barcelona, Spain}

\begin{document}

\section{Introduction}
The origin of the ultra-high-energy cosmic rays (UHECRs, $E>10^{18}$ eV = 1 EeV) is still not resolved. 
While considered extragalactic in origin \cite{Aab2017}, the astrophysical sources are unknown. 
Possible candidates include Active Galactic Nuclei (AGN) and their relativistic jets, starburst galaxies, 
gamma-ray bursts as well as cosmic (supercluster) shocks. The present paper focuses on the relevance
of the former, and highlights some of the recent developments in the field.\\
Experimentally, major progress has been achieved over the last couple of years by the Pierre Auger 
(PA) and the Telescope Array (TA) collaborations, see e.g. refs.~\cite{Castellina2019,Ogio2019} for recent 
reports. Both collaborations run hybrid instruments that are fully operational since 2008. The larger PA 
observatory consists of 1660 surface and four fluorescence detectors, and is located in the southern 
hemisphere (Mendoza, Argentina). The TA, on the other hand, consists of 507 surface and three 
fluorescence detectors, and is located in the northern hemisphere (Utah, USA). The surface detectors 
measure air shower particles on the ground, and are sensitive to its electromagnetic, muonic and hadronic 
components, while the fluorescence detectors observe the longitudinal development of air showers in 
the atmosphere by the light emitted during their passage.\\
In the following, basic experimental findings are briefly summarized. Helpful overviews of recent results 
and progress can also be found in refs.~\cite{Kachelriess2019,Anchordoqui2019}.

\section{Basic Experimental Results}
\subsection{Energy Spectrum}\label{sec_spectrum}
\begin{figure}[htbp]
\begin{center}
\includegraphics[width = 0.98 \textwidth]{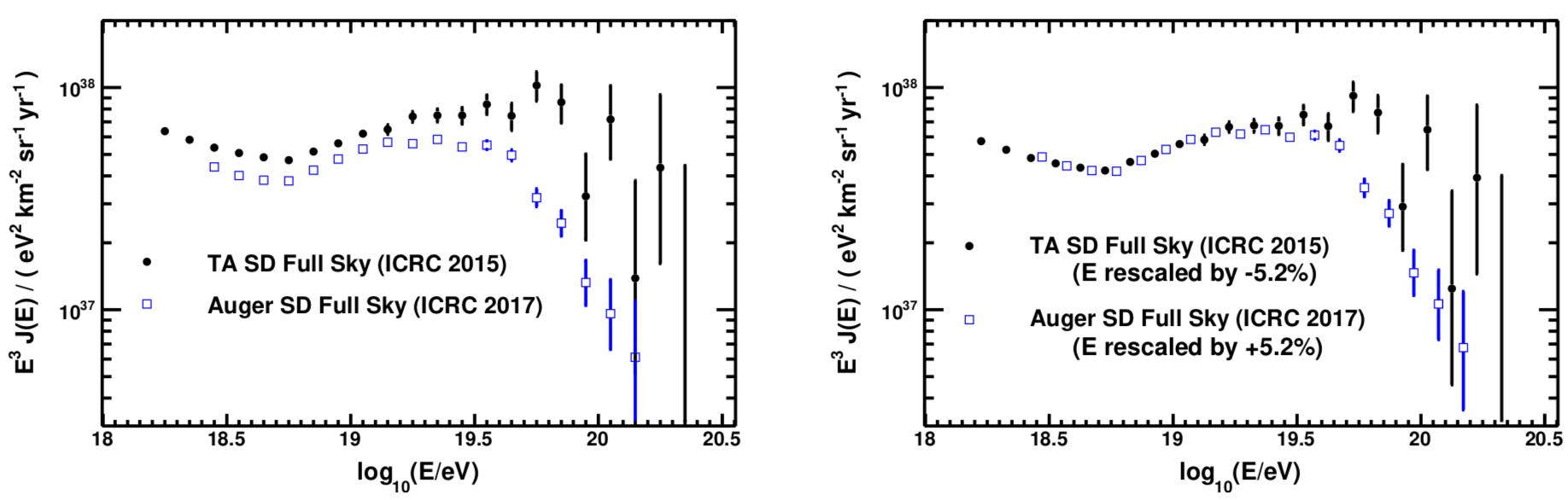}
\caption{Energy spectrum as measured by the PA and TA surface detectors (left), adjusted for uncertainties 
in absolute energy scale (right). From ref.~\cite{Ivanov2017}.}
\label{spectrum}
\end{center}
\end{figure}

A decent agreement between the TA and PA experiments is found up to $E\sim 4\times 10^{19}$ eV, 
once the data are adjusted for the uncertainty ($10\%$) in absolute energy scale, see e.g. ref.~\cite{Ivanov2017}
and Fig.~\ref{spectrum}. 
The resultant cosmic-ray (CR) energy spectrum also reveals a pronounced steepening above $E\sim 5\times 
10^{19}$ eV, that is compatible with a Greisen-Zatsepin-Kuzmin (GZK) cut-off. Differences between experiments 
are seen, however, towards higher energies. It seems possible that these may be caused by different instrumental 
systematics, and/or some true source difference (as the particle mean free path increases, differences in the 
large-scale structure may no longer be negligible).
The observed spectrum implies a UHECR luminosity density (around $10^{19.5}$ eV) of $l_{\rm UHECR} 
\sim 6\times 10^{43}$ erg/(Mpc$^3$ yr) \cite{Murase2019}.

\subsection{Composition}\label{sec_composition}
\begin{figure}[htbp]
\begin{center}
\includegraphics[width = 0.90 \textwidth]{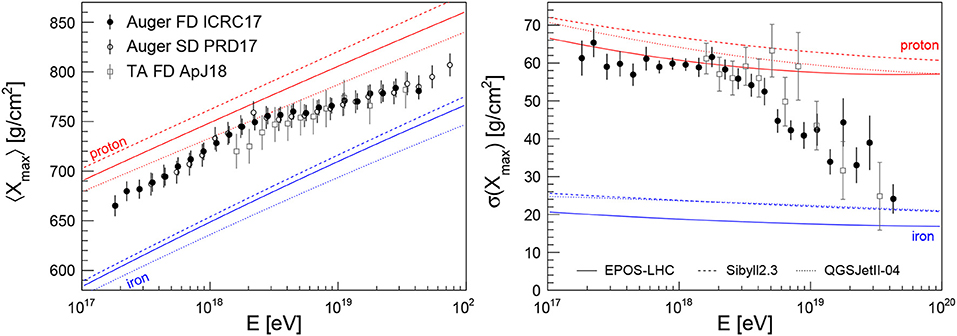}
\caption{Composition analysis based on the mean (left) and the standard deviation (right) of the distribution
of shower maximum as function of energy. The TA data have been corrected for detector effects and 
energy uncertainty. From ref.~\cite{Alves_Batista2019}.}
\label{composition}
\end{center}
\end{figure}

Cosmic-ray particles entering the atmosphere induce hadronic showers by interacting with atmospheric 
nuclei. The atmospheric depth of the shower maximum, $X_{\rm max}$ [g/cm$^2$], is sensitive to the 
primary mass. In general, light particles penetrate deeper in the atmosphere and exhibit a steeper lateral 
distribution compared to heavy nuclei. Comparison of TA and PA data suggests that the composition 
becomes lighter between $10^{17.2}$ eV and $10^{18.3}$ eV, compatible with a transition from galactic to 
extragalactic cosmic rays. Above $E\sim 10^{18.3}$ eV the composition seems to become heavier again, 
as evident both, from the mean $X_{\max}$ and its fluctuations $\sigma(X_{max})$ measurements (see 
Fig.~\ref{composition}), with a trend that protons are gradually replaced by helium, helium by nitrogen etc, 
an iron contribution possibly emerging above $10^{19.4}$ eV, cf. ref.~\cite{Alves_Batista2019} for details.

\subsection{Anisotropy on Intermediate Scale}\label{sec_anisotropy}
\begin{figure}[htbp]
\begin{center}
\includegraphics[width = 0.65 \textwidth]{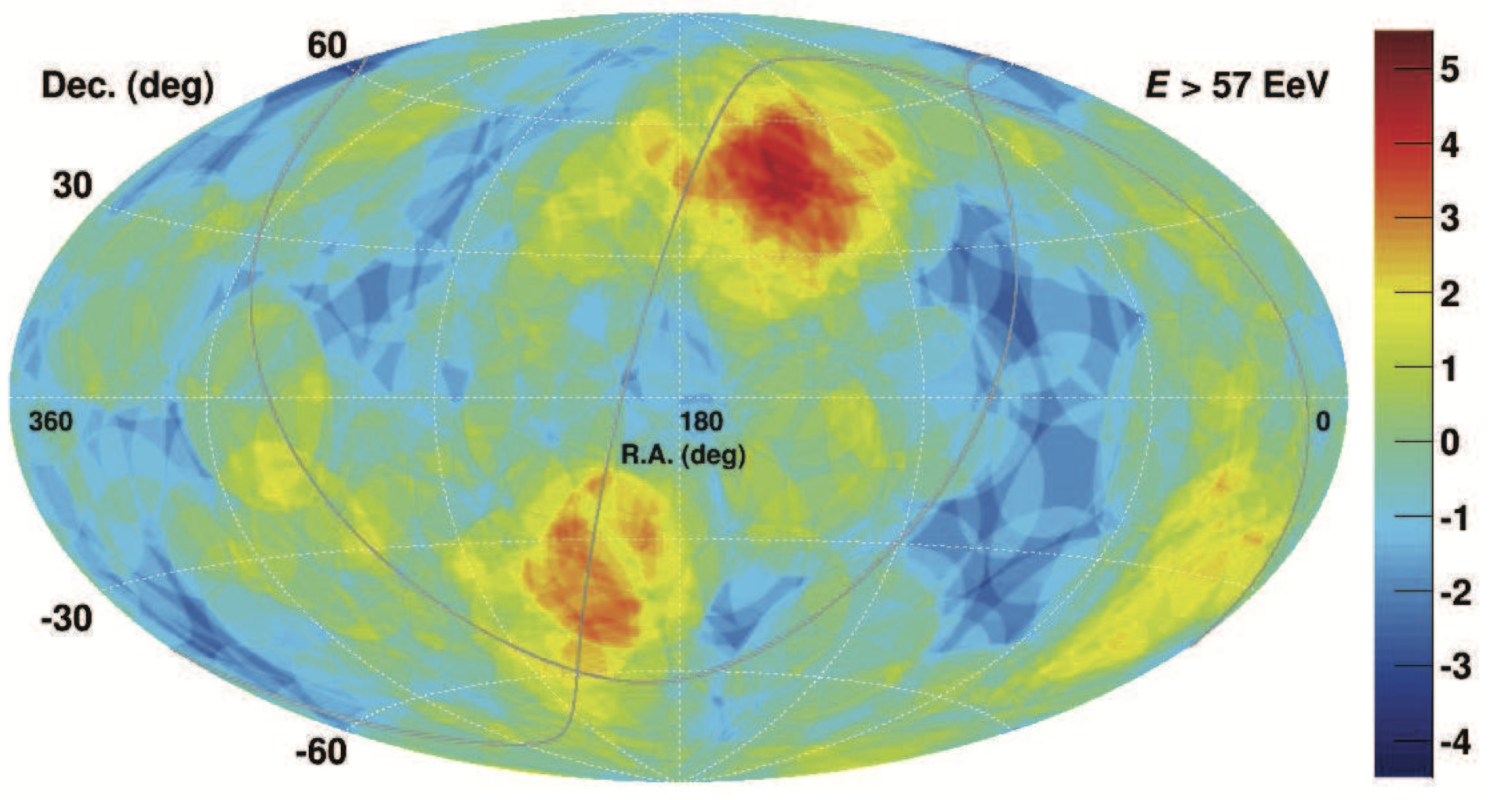}
\caption{Sky map (Hammer-Aitoff projection in equatorial coordinates) combining TA (109 events) 
and PA (157 events) data with energy above $5.7\times10^{19}$ eV. The data have oversampling 
with a $20^{\circ}$ radius circle. No energy corrections was applied. The thin gray line above and 
left of the TA (upper) "hot" spot is the supergalactic plane. The PA data also suggest a (lower) 
"warm" spot. From ref.~\cite{Matthews2017}.}
\label{anisotropy}
\end{center}
\end{figure}

A possible anisotropy in the arrival directions of UHECRs can offer important clues as to their astrophysical 
origin. A recent TA analysis \cite{Matthews2017} of the anisotropy on intermediate angular scales using TA 
and PA events above $5.7\times 10^{19}$ eV (smeared out on circles of $25^{\circ}$) provides evidence 
for a TA hot spot (at R.A. $\sim144^{\circ}$, dec $\sim40^{\circ}$) at a level of $5.2 \sigma$ (local significance) 
and $3.4 \sigma$ (global significance), respectively (cf. also \cite{Kawata2019}). This TA hot spot lies 
approximately in the direction of the Ursa Major/Virgo supercluster ($d \sim18$ Mpc). The PA data, on the 
other hand, also suggests the existence of a PA "warm" spot (local significance $3.6\sigma$), that coincides 
with the direction to Cen A (and the Centaurus supercluster at $d\sim50$ Mpc), see Fig.~\ref{anisotropy}.
Neither the TA nor the PA data shows any sign of excess in the direction of Virgo.\\ 
Note that if the UHECR composition would indeed become heavier ($Z\geq10$), as suggested in 
Sec.~\ref{sec_composition}, the standard requirements on astrophysical accelerators become less 
dramatic. 
On the other hand, since particle deflection in a regular magnetic fields scales with $\theta \sim d/r_g 
\propto (Z/E)$, one would then need to address why we do not observe a strong anisotropy associated 
with protons at $E/Z$ \cite{Lemoine2009,Liu2013,Lemoine2018}.

\subsection{Correlations with Known Sources}
The current PA analysis of the arrival direction of UHECRs above $20$ EeV ($\sim 5500$ events) appears to 
favour a correlation with starburst galaxies \cite{Aab2018}: In particular, a starburst model that attributes 
$9.7\%$ of UHECRs($>39$ EeV) to nearby starburst galaxies (23 objects, including NGC 4945, NGC 253, 
M83, NGC 1068), and the remaining to isotropic background, is favoured at $4\sigma$ over the hypothesis of 
isotropy, cf. also ref.~\cite{Caccianiga2019} for update, but see also ref.~\cite{Abbasi2018} for a (negative) TA test. 
In comparison, a model of nearby $\gamma$-bright AGN (17 objects, including Cen A, M87, Mkn 421, Mkn 501, 
but not Fornax A), which attributes $\sim7\%$ of the total flux ($>60$ EeV) to them, is (only) favoured against 
isotropy at $\sim2.7\sigma$. It has been suggested, however, that an incorporation of Fornax A (at a 
distance $\sim 20$ Mpc) could change this picture in favour of radio galaxies once magnetic deflection is 
properly accounted for \cite{Matthews2018,Matthews2019b}. 
The analysis is certainly further complicated by the fact that the CR luminosity of individual AGN is not known. 
In the noted PA analysis \cite{Aab2018}, the (Fermi-LAT) integral gamma-ray fluxes from 50 GeV to 2 TeV have 
been used as proxy for the UHECR flux. In the AGN case, the observable UHECR flux is then dominated 
($75\%$) by the core of Cen~A. The limitations introduced by the chosen approach necessitate further studies, 
part of which should include a basic treatment of magnetic deflection and the introduction of better flux proxies. 
As things are, it is still too early to conclude about the astrophysical sources of UHECRs.

In general, since UHECRs in AGN are confined to magnetic fields, they may escape only slowly from their 
sources. Since it seems likely that, e.g., Cen~A and Fornax A have seen more powerful jets in the past, capable 
of accelerating UHECRs (see below), UHECR particles could still be escaping from their giant lobes. If this is the 
case, then their past activity aka source history becomes relevant. For some exemplary, recent discussion of 
individual source associations the reader is referred to refs.~\cite{Eichmann2018} (Cen A), \cite{Kobzar2019} 
(Virgo A/M87), \cite{Fraija2019} (Cen B) and \cite{Matthews2018} (Fornax A), respectively.

\section{Physics Constraints}\label{constraints}
In order for a source to be capable of accelerating UHECRs, the relevant particles need to be confined within it. 
This introduces a general (Hillas) bound
\begin{equation}\label{Hillas}
E \leq 10^{20} \,Z\, (B/1\mu\mathrm{G}) (L/100~\mathrm{kpc})\;\mathrm{eV}\,,
\end{equation} given by the condition that the particle gyro-radius, $r_{\rm gyro}$, remains smaller than the 
characteristic dimension $L$ of the source. Similarly, assuming acceleration to be constrained by the motional 
electric field $\vec{\epsilon}= - (\vec{V}/c) \times \vec{B}$, one obtains
\begin{equation}\label{Hillas_2}
E \leq Z e \epsilon L = 10^{20}\, Z\, \beta \,(B/1\mu\mathrm{G}) (L/100~\mathrm{kpc})\;\mathrm{eV}\,,
\end{equation} with $\beta =V/c$. Accordingly, UHECR acceleration generally requires fast speeds, strong 
magnetic fields or large source volumes \cite{Hillas1984}. The requirements are significantly relaxed if the 
composition at the highest energies would be heavy ($Z\geq 10$). The above expressions, however, only 
provides a necessary (and not itself sufficient) condition for UHECR acceleration, and also neglect possible 
relativistic effects. Generalisations and improved constraints have been obtained by various considerations,
e.g. \cite{Norman1995,Blandford2000,Aharonian2002,Lemoine2009}. In any case, the characteristic timescale 
for acceleration $t_{\rm acc}$ has to be smaller than the escape ($\tau_{\rm esc}$) and the radiative ($\tau_{\rm 
loss}$) loss timescale, respectively. Note that since $t_{\rm acc}$ depends on acceleration physics, and 
$\tau_{\rm esc}$, $\tau_{\rm loss}$ on individual source physics, this also implies that detailed 
("multi-messenger") source studies become particularly interesting.\\
%
Considering shock-type particle acceleration in a relativistic outflow (of speed $\beta=V/c$ and Lorentz factor 
$\Gamma$), a generalized constraint \cite{Lemoine2009} is obtained by requiring that in the local, co-moving 
frame $t_{\rm acc}' < t_{\rm dyn}'$ (with primed quantities referring to the co-moving frame). Expressing the acceleration 
timescale in terms of the gyro-time, i.e. $t_{\rm acc}' =\eta t_{\rm gyro}' = \eta E'/(Z eB'c)$, $\eta\geq 1$, and 
the dynamical timescale as $t_{\rm dyn}' = d/(\Gamma\,V)$ (with longitudinal length scale $d$), the maximum 
UHECR energy becomes
\begin{equation}\label{gHillas}
E = \Gamma E' \leq Z e B' d /(\beta \eta)\,.
\end{equation} To allow for this, the magnetic luminosity of the source, $L_B = 2\pi r^2 \Gamma^2 u_B' V$ ($r$
the lateral half width, $u_B' \equiv B'^2/8\pi$), has to be high enough. With $B'$ from eq.~(\ref{gHillas}) and 
$r \sim \theta_j d$ ($\theta_j$ the jet half opening angle), one obtains $L_B \geq \theta_j^2 \Gamma^2 \eta^2 
E^2 \beta^3 c/(2\, Z\, e)^2$, or
\begin{equation}\label{lower_limit}
L_B \gppr 8 \times 10^{44} \theta_j^2 \Gamma^2 \eta^2 \beta^3 \left(\frac{E/Z}{10^{20}~\mathrm{eV}}\right)^2
                    \, \mathrm{erg/sec}\,. 
\end{equation} For the commonly assumed $\theta_j \sim 1/\Gamma$, this thus implies a lower limit ($\eta =1$)
on the required source luminosity for steady relativistic ($\beta \sim 1$) outflows of $L_m \gppr 8\times 10^{44} 
([E/Z]/10^{20}\mathrm{eV})^2$ erg/sec. Note that for non-relativistic shocks, $\eta \sim (c/V_s)^2$, so that $L_m 
\propto (c/V_s)$. The situation is again much relaxed if a heavier composition prevails at the highest end. Note
that while AGN can have fast jets with maximum power reaching up to $L_j \sim 10^{48} \dot{m}\,(M_{BH}/10^9
\,M_{\odot})$ erg/sec \cite{Katsoulakos2018}, the winds in starburst galaxies usually only have $L_{SBG} \sim 
10^{42}$ erg/sec (along with low shock velocities $\sim 1000$ km/s) \cite{Romero2018,Matthews2018}. 
Taken as face value, this would disfavour starburst galaxies as promising UHECR accelerators.

\section{UHECR Acceleration in AGN}
UHECR acceleration in AGN may occur at various locations, close to the black hole, within their jets, at 
hot spot shocks or in their large-scale lobes. In the following, three recent examples will be briefly discussed. 

\subsection{Black Hole Vicinity}
\begin{figure}[htbp]
\begin{center}
\includegraphics[width = 0.65 \textwidth]{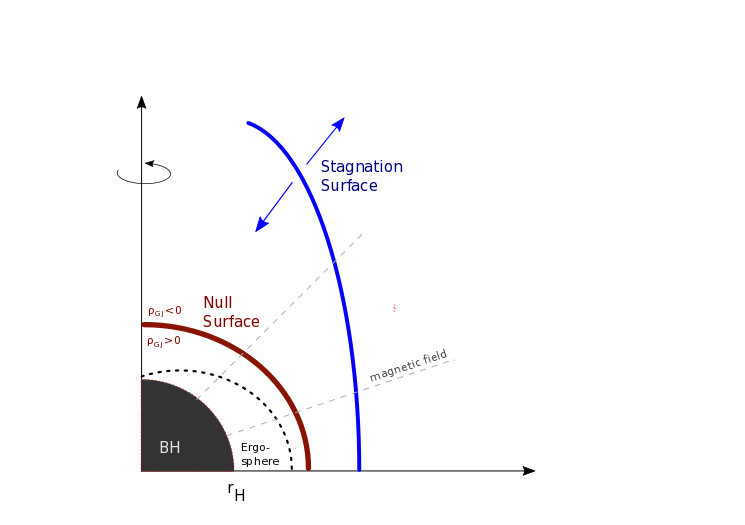}
\caption{Illustration of the possible locations of charge-deficient regions (gaps) in rotating black hole 
magnetospheres where efficient UHECR acceleration may occur. The red line denotes the null surface 
across which the required Goldreich-Julian charge density changes sign, while the blue line delineates 
the stagnation surface from which stationary MHD flows start.}
\label{gap}
\end{center}
\end{figure}

UHECR acceleration in the magnetospheres of rotating supermassive black holes has been considered in
a number of papers, see e.g. refs.~\cite{Levinson2000,Neronov2009,Rieger2011,Ptitsyna2016,Moncada2017}.
The acceleration relies on the occurrence of an electric field component parallel to the magnetic field in a 
charge-deficient ("gap") region close to the black hole, see also ref.~\cite{Rieger2018} for a review. This 
could happen either at the so-called null surface (across which the Goldreich-Julian charge density, required 
to screen the field, changes sign) or at the stagnation surface (separating MHD in- and outflows), see 
Fig.~\ref{gap}.

The maximum available voltage drop for a gap of width $h$ is of the order of \cite{Katsoulakos2018} 
\begin{equation}\label{voltage_drop}
\Delta \Phi = \frac{1}{c} \Omega_F r_H^2 B_H \left(\frac{h}{r_g}\right)^2 
                     \simeq 2\times 10^{21} \dot{m}^{1/2} M_9^{1/2} \left(\frac{h}{r_g}\right)^2\,,
\end{equation} assuming a magnetic field $B_H \simeq 2\times 10^5 \dot{m}^{1/2} M_9^{-1/2}$ G, 
with $M_9 = M_{\rm BH}/10^9 M_{\odot}$, field line rotation $\Omega_F= \Omega_H/2=c/4r_g$ and 
gravitational radius $r_g=GM/c^2$. This would seem to suggest that in massive ($M_9 \gppr 1$), 
active ($\dot{m}\gppr 10^{-3}$) sources, ultra-high energies $Z\,e\,\Delta\Phi \sim 10^{20}Z$ eV might 
well be achievable.\\ 
Under realistic conditions, however, energy losses due to curvature radiation are likely to introduce an 
upper limit \cite{Katsoulakos2018}
\begin{equation}
 \gamma_{\rm max} \simeq 10^{10} \frac{\dot{m}^{1/8}}{Z^{1/4}} M_9^{3/8} \left(\frac{h}{r_g}\right)^{1/4}
\end{equation} for curvature radii of the order of the gravitational one. Note that this would imply a 
maximum energy $E_{\rm max}$ which no longer scales linearly with $Z$. Moreover, unless the accretion 
rate is low enough (usually, $\dot{m} \leq 10^{-4}$), the ambient soft photon environment will facilitate 
efficient pair production in the gap, leading to gap sizes $h \ll r_g$, significantly reducing achievable 
particle energies (eq.~[\ref{voltage_drop}]). On the other hand, if accretion rates are too low (as in e.g. 
quiescent quasars), the expected magnetic field strength ($B_H \propto \dot{m}^{1/2}$) would be low 
as well.\\ 
When taken together, this would seem to make an efficient gap-type acceleration of protons to energies 
much beyond $E \sim 10^{18}$ eV problematic. We note that since efficient UHECR production in the 
black hole magnetosphere is accompanied by curvature VHE emission, gamma-ray observations may 
in principle allow a useful probe of UHECR acceleration, e.g. \cite{Levinson2000,Pedaletti2011}.

\subsection{Shear in large-scale Jets}
The jets in AGN are likely to exhibit some internal jet stratification and velocity shearing, that could be 
conducive to efficient cosmic-ray acceleration, see ref.~\cite{Rieger2019} for review and discussion. 
Prominent scenarios include Fermi-type particle acceleration in non-gradual (discontinuous), e.g. refs.
\cite{Ostrowski1998,Ostrowski2000,Caprioli2015,Kimura2018}, or gradual (continuous), e.g. refs. \cite{Rieger2004,
Rieger2016,Liu2017,Webb2018,Webb2019}, velocity shear flows.

\begin{figure}[htb]
\begin{center}
\includegraphics[width=0.48 \textwidth]{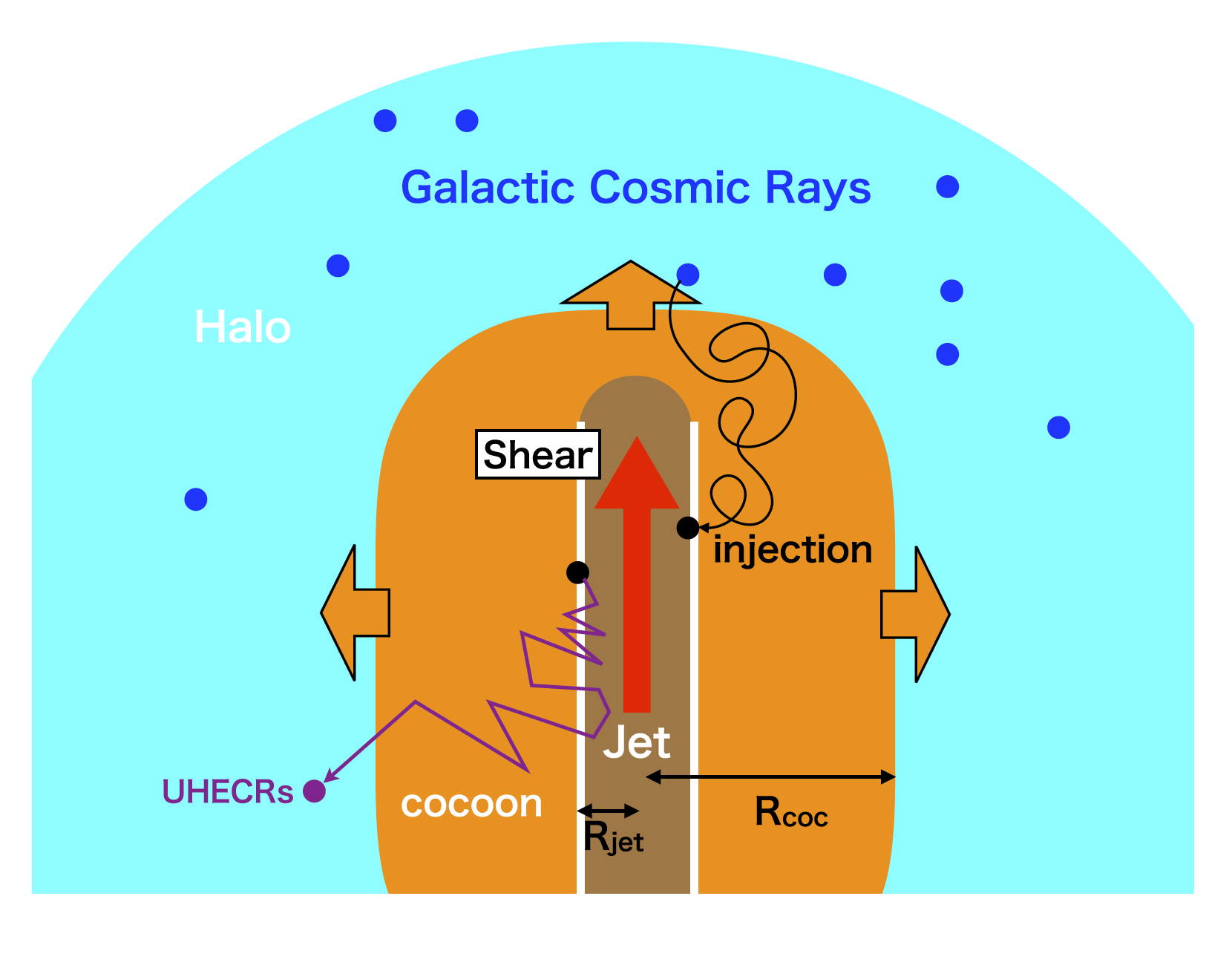}
\includegraphics[width=0.47 \textwidth]{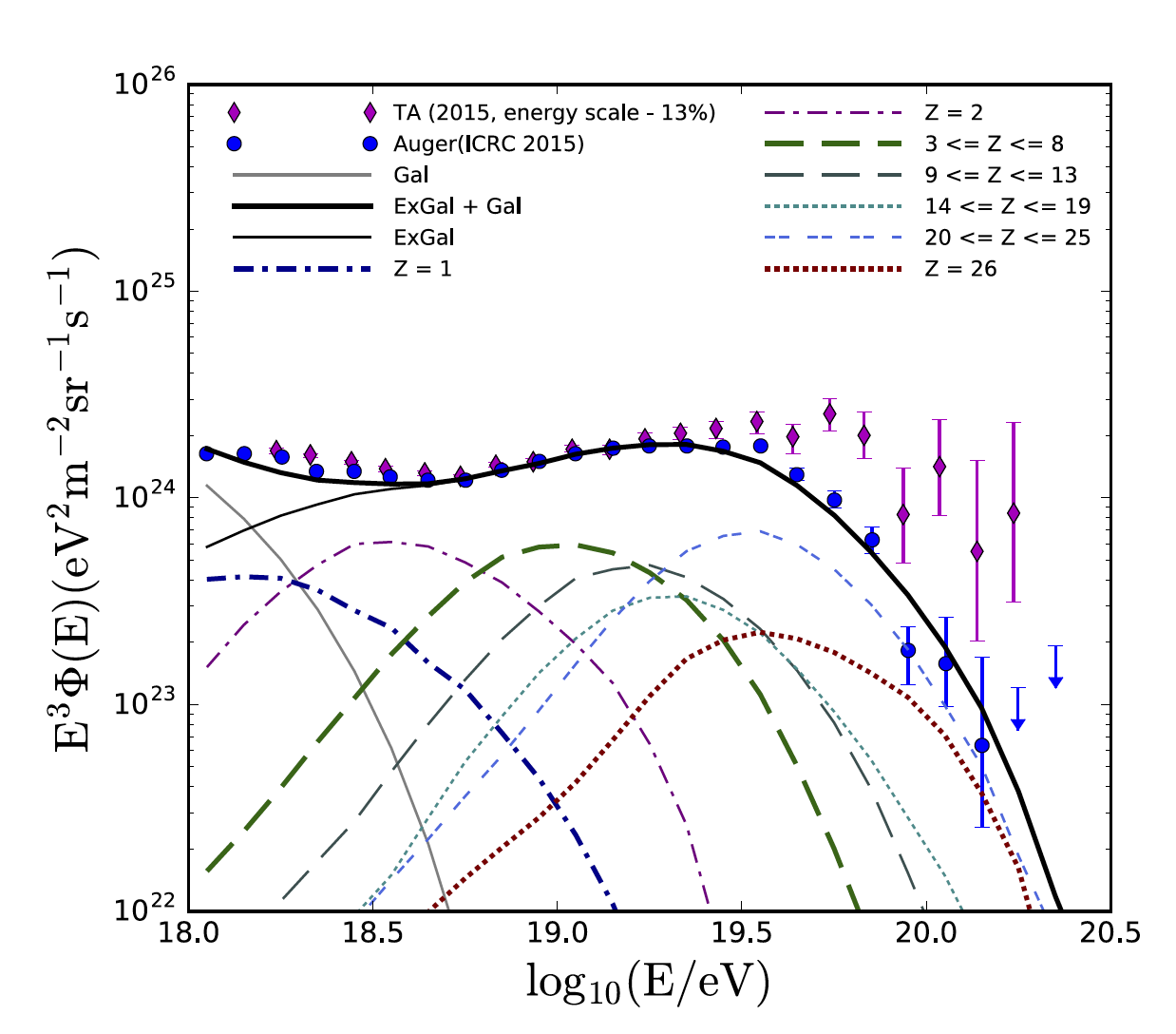}
\caption{{\bf Left:} Sketch of the considered FR~I model assuming a recycling of galactic cosmic rays by 
non-gradual shear particle acceleration in a jet - (turbulent) cocoon system. Some fraction of galactic cosmic 
rays is considered to be swept up ("injected") and reaccelerated to ultra-high energies. The return probability 
(and thus the cycle time) of a particle is dominated by the scattering (turbulence) properties in the cocoon. 
{\bf Right:} Reconstruction of the observed UHECR spectrum assuming a mildly relativistic ($\beta \simeq 0.7$) 
jet of width $r_j=0.5$ kpc surrounded by a thin velocity transition layer $\Delta r=5$ pc ($B_j=0.3$mG). The 
chemical composition at the highest energies is dominated by intermediate and heavy nuclei. From 
ref.~\cite{Kimura2018}.}
\label{Kimura}
\end{center}
\end{figure}

The former utilises that if the CR particle distribution would remain nearly isotropic near a strong shear discontinuity, 
the mean fractional energy change for crossing and re-crossing (full cycle) is given by $\left< \Delta E/E \right> 
\simeq \Gamma^2 \beta^2$ \cite{Rieger2004}. This suggests that the increase in particle energy could be substantial 
for a velocity shear that is highly relativistic ($\Gamma \gg 1$, e.g. \cite{Caprioli2015}), while for non-relativistic 
speeds ($\Gamma \sim 1$) only the usual gain $\propto \beta^2$ is obtained. For relativistic flow speeds ($\beta 
\simeq 1$), however, the non-negligible anisotropy of the particle distribution has to be modelled and taken into 
account. For repeated crossings, the principal effects of this is a reduction in efficiency, e.g. \cite{Ostrowski1998,
Ostrowski2000}.\\ 
Denoting by $\tau$ the mean cycle time (into and out of the shear), the mean acceleration timescale might in 
general be expressed as $t_{\rm acc} \simeq \tau / \left<\Delta E/E \right>$. Note that, depending 
on the considered turbulence properties, $\tau$ might actually be dominated by the (diffusion) time a CR particle 
needs to return to the jet shear \cite{Kimura2018}. 
A related application to the kiloparsec-scale jets of Fanaroff-Riley (FR) I sources has been recently presented 
\cite{Kimura2018}, see Fig.~\ref{Kimura}. The model assumes entrainment and re-acceleration of Galactic 
cosmic rays in a mildly relativistic shear flow ($\beta = 0.7$). The performed Monte Carlo simulations suggest 
that the escaping CRs can have quite hard spectra ($dN/dE \propto E^{-a}, a \lppr 1$). Furthermore, a rather 
complex chemical compositions at UHECR is achieved due to different injection at TeV-PeV energies 
($E_{\rm inj,i} =15\,Z_i$ TeV). This also allows to accommodate the anisotropy constraints mentioned earlier 
(Sec.~\ref{sec_anisotropy}). The maximum energy $\propto Z$ is limited by the jet size ($t_{\rm acc} \propto 
r_j/c$) and (via $\tau$) dominated by the diffusion properties in the cocoon. 
The results shown are sensitive to the chosen cocoon properties (i.e., cocoon size, turbulence scale) and 
dependent on a thin velocity transition layer $\Delta r \sim r_j/100$ (defining the required energy of the 
injected galactic cosmic  rays). Enlarging $\Delta r$, for example, is likely to affect the outcome. 
Nevertheless, these simulations show that non-gradual shear acceleration in large-scale jets of FR~I (and
not only FR~II) could in principle play an important role in UHECR acceleration. Given an FR~I number 
density of $n_{FRI} \sim10^{-5}-10^{-4}$ Mpc$^{-3}$, an average source luminosity $L\sim 2\times 10^{40}-
2\times 10^{41}$ erg/sec would be needed, cf. Sec.~\ref{sec_spectrum}.

\begin{figure}[hbt]
\begin{center}
\includegraphics[width=0.50 \textwidth]{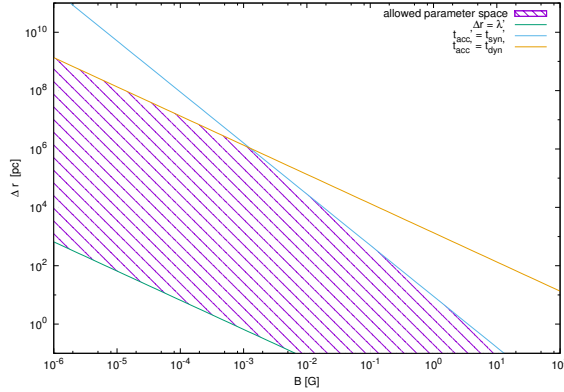}
\caption{Permitted (hatched) parameter range (magnetic field strength $B$, shear layer width $\Delta r$) to 
allow gradual shear acceleration of protons to $\sim10^{18}$ eV and to satisfy confinement and synchrotron 
loss constraints. A Kolmogorov-type scaling for the particle mean free path ($q=2-\alpha=5/3$) has been 
assumed. The required conditions for UHE proton acceleration might be met in the large-scale jets of AGN.}
\label{Liu2}
\end{center}
\end{figure}

If the transition layer becomes larger, particle scattering will occur within the shear layer, facilitating {\it gradual 
shear particle acceleration} within it, see e.g. refs.~\cite{Rieger2004,Rieger2016,Liu2017,Webb2018,Webb2019}. 
The underlying mechanism can be viewed as a stochastic, second-order Fermi-type particle acceleration process, 
where the usual scattering center speed is replaced by an effective velocity $\bar{u}$ determined by the 
shear flow profile, e.g., $\bar{u} = (\partial u_z/\partial r)\, \lambda$ in the case of a simple continuous 
(non-relativistic) velocity shear $\vec{u} = u_z(r) \vec{e}_z$, cf.  ref.~\cite{Rieger2019} for a recent review.
Accordingly, the fractional energy change scales as 
\begin{equation}
\left< \Delta E/E \right> \propto \left(\frac{\bar{u}}{c}\right)^2 
                                                                    \propto \left( \frac{\partial u_z}{\partial r} \right)^2 
                                                                    \lambda^2\,.
\end{equation} This suggests a scaling for the characteristic acceleration timescale of $t_{\rm acc} \simeq 
\tau / \left<\Delta E/E \right> \propto 1/\lambda$, which, in contrast to classical first-order Fermi (shock) 
as well as non-gradual shear acceleration, is inversely proportional to the particle mean free path $\lambda 
= c\tau$. This seemingly unusual behaviour relates to the fact that as a particle increases its energy ($E 
\simeq p c $), and thereby its mean free path (typically, $\lambda(p) \propto p^{\alpha}, \alpha >0$), a 
higher effective velocity $\bar{u}$ is experienced.\\ 
Gradual shear particle acceleration is particularly interesting as it could offer an explanation for the origin 
of the extended synchrotron X-ray emission in large-scale AGN jets that requires the maintenance of 
ultra-relativistic electrons ($\gamma_e \sim 10^8$) on kpc-scales \cite{Liu2017}.
Efficient operation generally requires sufficiently relativistic flow speeds, i.e. fast jets 
(bulk Lorentz factors of some few) or a strong jet--back-flow system \cite{Webb2018,Webb2019,Rieger2019}.\\
To assess its potential for UHECR acceleration, cf. Sec.~\ref{constraints}, the properties of a source need to 
be such as to allow (i) CRs to be confined within its jet, and to enable CR acceleration (ii) to proceed faster 
than radiative losses ($t_{\rm acc} <t_{\rm syn}$) and (iii) to operate within the lifetime ($t_{\rm dyn}$) of the system. A related
application following ref.~\cite{Liu2017} is shown in Fig.~\ref{Liu2} assuming a linearly decreasing flow profile
with $\Gamma=2$ on the jet axis. A jet-width to length ratio $\Delta r/d=0.02$ has been employed.
These results suggest that proton energies $\sim10^{18}$ eV are in principle achievable for relatively plausible 
parameters (e.g., jet lengths 10 kpc -1 Mpc, magnetic fields $B \sim [1-100]~\mu$G). Higher energies might be 
obtained for faster flows or heavier particles \cite{Liu2017,Webb2018,Webb2019}.

\subsection{Multiple shocks in back-flows}
\begin{figure}[hbt]
\begin{center}
\includegraphics[width=0.30 \textwidth]{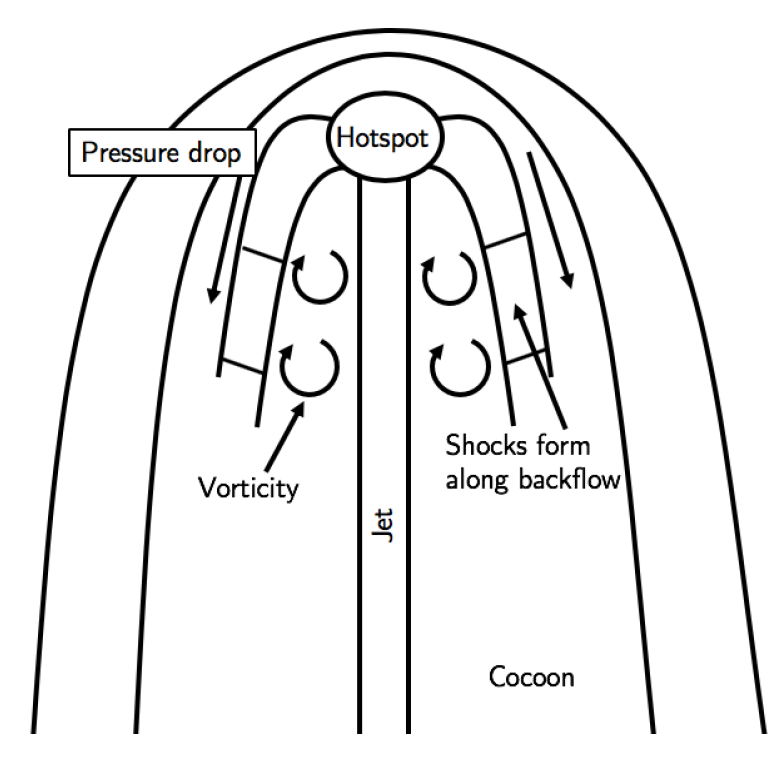}
\caption{Cartoon of the considered scenario, where multiple (mildly relativistic) shocks in strong back-flows 
provide sites for efficient CR acceleration to UHE energies. From ref.~\cite{Matthews2019b}.}
\label{Matthews}
\end{center}
\end{figure}

Following the considerations in Sec.~\ref{constraints} and eq.~(\ref{Hillas_2}), high speeds are seemingly
conducive for efficient UHECR acceleration. While highly relativistic shocks might thus appear most promising, 
closer studies however reveal them to be problematic instead, see e.g. refs~\cite{Lemoine2010,Sironi2011,
Bell2018}. This is partly related to the fact that highly relativistic shocks are generically perpendicular (with 
downstream magnetic field quasi perpendicular to the shock normal, preventing particle from diffusing back 
upstream) and that particle isotropization upstream is no longer guaranteed (there being not sufficient time to 
growth turbulence on scales $r_{\rm gyro,UHECR}$). The situation is relaxed for mildly relativistic shock 
speeds, and this has led to the recent proposal that UHECR particle acceleration may instead occur at 
multiple (mildly relativistic) shocks in the back-flows of radio galaxies \cite{Matthews2019}. Multiple shocks
would provide multiple opportunities for acceleration, and also lead to harder spectra.\\
The proposal is motivated by two- and three-dimensional hydrodynamical simulations of light (density contrast
$\rho_j/\rho_0 \sim 10^{-5}-10^{-4}$), high-power ($\sim 10^{45}$ erg/sec) jets in cluster environments, in which
strong back-flows are seen (cf. also \cite{Perucho2007,Perucho2019}), that can be supersonic and exhibit related 
compression structures, cf. Fig.~\ref{Matthews}.
Analysis of the simulation data suggests that about $10\%$ of the particles pass through a shock of Mach 
number $M>3$, while $\sim5\%$ pass through multiple shocks. Inferred shock speeds are of the order of 
$u_s\sim0.2$ c, with estimated sizes $R \sim2$ kpc. Assuming a reference magnetic field strength of $B\sim 
100\,\mu$G, maximum CR energies of $E_{\rm max} \simeq 4 \times 10^{19} Z \,
(B/10^{-4}\mathrm{G})~(R/2\mathrm{kpc})$ eV, see eq.~(\ref{Hillas_2}), could thus be achievable. While
radio galaxies can in principle be powerful enough to satisfy the minimum power requirement $L_m$, 
eq.~(\ref{lower_limit}), this is currently barely the case within the GZK horizon. As CR are likely to escape
slowly from a source, however, the past activity of a source would be relevant (see above). There are 
good reasons, for example, to consider an enhanced activity in the past driving the giant lobe evolution 
in Cen~A. In the present context these lobes may then represent reservoirs of UHECRs.

\section{Conclusion and Perspectives}
Current correlation studies provide some indications that both starburst galaxies and AGN could play a 
significant role in the production of UHECRs. From a physical (acceleration) point of view, starburst 
galaxies are less promising, while radio galaxies appear capable of satisfying the relevant efficiency 
requirements. In particular, Fermi-type particle acceleration at trans-relativistic (internal) shocks and/or 
in shearing, relativistic jet flows could facilitate the production of UHECRs in jetted AGNs. This could be 
compatible with recent findings suggesting an increased ($\sim 4\sigma$) excess of UHECR ($E>39$ 
eV) events around Cen A. Further experimental constraints on the chemical composition as well as 
more complex (and physically motivated) correlation studies will help to eventually conclude about the 
astrophysical sources of UHECRs.\\ 

\noindent
{\small Acknowledgement: I am grateful to the organisers for the invitation to a stimulating 
conference. I thank James Matthews and Kohta Murase for allowance to use figures from
their papers. Funding by a DFG Heisenberg Fellowship RI 1187/6-1 is kindly acknowledged.}

\bibliographystyle{JHEP}
\bibliography{paper_biblio.bib}

\providecommand{\href}[2]{#2}\begingroup\raggedright\begin{thebibliography}{10}

\bibitem{Aab2017}
{Pierre Auger Collaboration}, A.~{Aab}, P.~{Abreu}, M.~{Aglietta}, I.~A.
  {Samarai}, I.~F.~M. {Albuquerque} et~al., \emph{{Observation of a large-scale
  anisotropy in the arrival directions of cosmic rays above $8\times10^{18}$
  eV}}, \href{https://doi.org/10.1126/science.aan4338}{\emph{Science}
  {\bfseries 357} (2017) 1266}
  [\href{https://arxiv.org/abs/1709.07321}{{\ttfamily 1709.07321}}].

\bibitem{Castellina2019}
A.~{Castellina}, \emph{{Highlights from the Pierre Auger Observatory}},  in
  \emph{36th International Cosmic Ray Conference (ICRC2019)}, vol.~36 of
  \emph{International Cosmic Ray Conference}, p.~4, Jul, 2019,
  \href{https://arxiv.org/abs/1909.10791}{{\ttfamily 1909.10791}}.

\bibitem{Ogio2019}
S.~{Ogio}, \emph{{Highlights from the Telescope Array}},  in \emph{36th
  International Cosmic Ray Conference (ICRC2019)}, vol.~36 of
  \emph{International Cosmic Ray Conference}, p.~13, Jul, 2019.

\bibitem{Kachelriess2019}
M.~{Kachelriess} and D.~V. {Semikoz}, \emph{{Cosmic Ray Models}},
  {\emph{Progress in Particle and Nuclear Physics} (2019) arXiv:1904.08160}
  [\href{https://arxiv.org/abs/1904.08160}{{\ttfamily 1904.08160}}].

\bibitem{Anchordoqui2019}
L.~A. {Anchordoqui}, \emph{{Ultra-high-energy cosmic rays}},
  \href{https://doi.org/10.1016/j.physrep.2019.01.002}{\emph{Physics Reports}
  {\bfseries 801} (2019) 1} [\href{https://arxiv.org/abs/1807.09645}{{\ttfamily
  1807.09645}}].

\bibitem{Ivanov2017}
D.~{Ivanov}, {Pierre Auger Collaboration} and {Telescope Array Collaboration},
  \emph{{Report of the Telescope Array - Pierre Auger Observatory Working Group
  on Energy Spectrum}},  in \emph{35th International Cosmic Ray Conference
  (ICRC2017)}, vol.~301 of \emph{International Cosmic Ray Conference}, p.~498,
  Jan, 2017.

\bibitem{Murase2019}
K.~{Murase} and M.~{Fukugita}, \emph{{Energetics of high-energy cosmic
  radiations}}, \href{https://doi.org/10.1103/PhysRevD.99.063012}{\emph{Phys
  Rev D} {\bfseries 99} (2019) 063012}
  [\href{https://arxiv.org/abs/1806.04194}{{\ttfamily 1806.04194}}].

\bibitem{Alves_Batista2019}
R.~{Alves Batista}, J.~{Biteau}, M.~{Bustamante}, K.~{Dolag}, R.~{Engel},
  K.~{Fang} et~al., \emph{{Open questions in cosmic-ray research at ultrahigh
  energies}}, \href{https://doi.org/10.3389/fspas.2019.00023}{\emph{Frontiers
  in Astronomy and Space Sciences} {\bfseries 6} (2019) 23}
  [\href{https://arxiv.org/abs/1903.06714}{{\ttfamily 1903.06714}}].

\bibitem{Matthews2017}
J.~{Matthews} and {Telescope Array Collaboration}, \emph{{Highlights from the
  Telescope Array Experiment}},  in \emph{35th International Cosmic Ray
  Conference (ICRC2017)}, vol.~301 of \emph{International Cosmic Ray
  Conference}, p.~1096, Jan, 2017.

\bibitem{Kawata2019}
K.~{Kawata}, A.~{di Matteo}, T.~{Fujii}, D.~{Ivanov}, C.~C.~H. {Jui}, J.~P.
  {Lundquist} et~al., \emph{{TA Anisotropy Summary}},  in \emph{European
  Physical Journal Web of Conferences}, vol.~210 of \emph{European Physical
  Journal Web of Conferences}, p.~01004, Oct, 2019,
  \href{https://doi.org/10.1051/epjconf/201921001004}{DOI}.

\bibitem{Lemoine2009}
M.~{Lemoine} and E.~{Waxman}, \emph{{Anisotropy vs chemical composition at
  ultra-high energies}},
  \href{https://doi.org/10.1088/1475-7516/2009/11/009}{\emph{JCAP} {\bfseries
  2009} (2009) 009} [\href{https://arxiv.org/abs/0907.1354}{{\ttfamily
  0907.1354}}].

\bibitem{Liu2013}
R.-Y. {Liu}, A.~M. {Taylor}, M.~{Lemoine}, X.-Y. {Wang} and E.~{Waxman},
  \emph{{Constraints on the Source of Ultra-high-energy Cosmic Rays Using
  Anisotropy versus Chemical Composition}},
  \href{https://doi.org/10.1088/0004-637X/776/2/88}{\emph{ApJ} {\bfseries 776}
  (2013) 88} [\href{https://arxiv.org/abs/1308.5699}{{\ttfamily 1308.5699}}].

\bibitem{Lemoine2018}
M.~{Lemoine}, \emph{{On Ultra-High Rigidity Cosmic Rays}},  in \emph{Ultra-High
  Energy Cosmic Rays (UHECR2016)}, p.~011004, Jan, 2018,
  \href{https://doi.org/10.7566/JPSCP.19.011004}{DOI}.

\bibitem{Aab2018}
A.~{Aab}, P.~{Abreu}, M.~{Aglietta}, I.~F.~M. {Albuquerque}, I.~{Allekotte},
  A.~{Almela} et~al., \emph{{An Indication of Anisotropy in Arrival Directions
  of Ultra-high-energy Cosmic Rays through Comparison to the Flux Pattern of
  Extragalactic Gamma-Ray Sources}},
  \href{https://doi.org/10.3847/2041-8213/aaa66d}{\emph{ApJL} {\bfseries 853}
  (2018) L29} [\href{https://arxiv.org/abs/1801.06160}{{\ttfamily
  1801.06160}}].

\bibitem{Caccianiga2019}
L.~{Caccianiga}, \emph{{Anisotropies of the Highest Energy Cosmic-ray Events
  Recorded by the Pierre Auger Observatory in 15 years of Operation}},  in
  \emph{36th International Cosmic Ray Conference (ICRC2019)}, vol.~36 of
  \emph{International Cosmic Ray Conference}, p.~206, Jul, 2019.

\bibitem{Abbasi2018}
R.~U. {Abbasi}, M.~{Abe}, T.~{Abu-Zayyad}, M.~{Allen}, R.~{Azuma},
  E.~{Barcikowski} et~al., \emph{{Testing a Reported Correlation between
  Arrival Directions of Ultra-high-energy Cosmic Rays and a Flux Pattern from
  nearby Starburst Galaxies using Telescope Array Data}},
  \href{https://doi.org/10.3847/2041-8213/aaebf9}{\emph{ApJL} {\bfseries 867}
  (2018) L27} [\href{https://arxiv.org/abs/1809.01573}{{\ttfamily
  1809.01573}}].

\bibitem{Matthews2018}
J.~H. {Matthews}, A.~R. {Bell}, K.~M. {Blundell} and A.~T. {Araudo},
  \emph{{Fornax A, Centaurus A, and other radio galaxies as sources of
  ultrahigh energy cosmic rays}},
  \href{https://doi.org/10.1093/mnrasl/sly099}{\emph{MNRAS} {\bfseries 479}
  (2018) L76} [\href{https://arxiv.org/abs/1805.01902}{{\ttfamily
  1805.01902}}].

\bibitem{Matthews2019b}
J.~H. {Matthews}, A.~R. {Bell}, A.~T. {Araudo} and K.~M. {Blundell},
  \emph{{Cosmic ray acceleration to ultrahigh energy in radio galaxies}},  in
  \emph{European Physical Journal Web of Conferences}, vol.~210 of
  \emph{European Physical Journal Web of Conferences}, p.~04002, Oct, 2019,
  \href{https://arxiv.org/abs/1902.10382}{{\ttfamily 1902.10382}},
  \href{https://doi.org/10.1051/epjconf/201921004002}{DOI}.

\bibitem{Eichmann2018}
B.~{Eichmann}, J.~P. {Rachen}, L.~{Merten}, A.~{van Vliet} and J.~{Becker
  Tjus}, \emph{{Ultra-high-energy cosmic rays from radio galaxies}},
  \href{https://doi.org/10.1088/1475-7516/2018/02/036}{\emph{JCAP} {\bfseries
  2018} (2018) 036} [\href{https://arxiv.org/abs/1701.06792}{{\ttfamily
  1701.06792}}].

\bibitem{Kobzar2019}
O.~{Kobzar}, B.~{Hnatyk}, V.~{Marchenko} and O.~{Sushchov}, \emph{{Search for
  ultra high-energy cosmic rays from radiogalaxy Virgo A}},
  \href{https://doi.org/10.1093/mnras/stz094}{\emph{MNRAS} {\bfseries 484}
  (2019) 1790} [\href{https://arxiv.org/abs/1810.10294}{{\ttfamily
  1810.10294}}].

\bibitem{Fraija2019}
N.~{Fraija}, M.~{Araya}, A.~{Galv{\'a}n-G{\'a}mez} and J.~A. {de Diego},
  \emph{{Analysis of Fermi-LAT observations, UHECRs and neutrinos from the
  radio galaxy Centaurus B}},
  \href{https://doi.org/10.1088/1475-7516/2019/08/023}{\emph{JCAP} {\bfseries
  2019} (2019) 023} [\href{https://arxiv.org/abs/1811.01108}{{\ttfamily
  1811.01108}}].

\bibitem{Hillas1984}
A.~M. {Hillas}, \emph{{The Origin of Ultra-High-Energy Cosmic Rays}},
  \href{https://doi.org/10.1146/annurev.aa.22.090184.002233}{\emph{ARA\&A}
  {\bfseries 22} (1984) 425}.

\bibitem{Norman1995}
C.~A. {Norman}, D.~B. {Melrose} and A.~{Achterberg}, \emph{{The Origin of
  Cosmic Rays above $10^{18.5}$ eV}},
  \href{https://doi.org/10.1086/176465}{\emph{ApJ} {\bfseries 454} (1995) 60}.

\bibitem{Blandford2000}
R.~D. {Blandford}, \emph{{Acceleration of Ultra High Energy Cosmic Rays}},
  \href{https://doi.org/10.1238/Physica.Topical.085a00191}{\emph{Physica
  Scripta Volume T} {\bfseries 85} (2000) 191}
  [\href{https://arxiv.org/abs/astro-ph/9906026}{{\ttfamily
  astro-ph/9906026}}].

\bibitem{Aharonian2002}
F.~A. {Aharonian}, A.~A. {Belyanin}, E.~V. {Derishev}, V.~V. {Kocharovsky} and
  V.~V. {Kocharovsky}, \emph{{Constraints on the extremely high-energy cosmic
  ray accelerators from classical electrodynamics}},
  \href{https://doi.org/10.1103/PhysRevD.66.023005}{\emph{Phys Rev D}
  {\bfseries 66} (2002) 023005}
  [\href{https://arxiv.org/abs/astro-ph/0202229}{{\ttfamily
  astro-ph/0202229}}].

\bibitem{Katsoulakos2018}
G.~{Katsoulakos} and F.~M. {Rieger}, \emph{{Magnetospheric Gamma-Ray Emission
  in Active Galactic Nuclei}},
  \href{https://doi.org/10.3847/1538-4357/aaa003}{\emph{ApJ} {\bfseries 852}
  (2018) 112} [\href{https://arxiv.org/abs/1712.04203}{{\ttfamily
  1712.04203}}].

\bibitem{Romero2018}
G.~E. {Romero}, A.~L. {M{\"u}ller} and M.~{Roth}, \emph{{Particle acceleration
  in the superwinds of starburst galaxies}},
  \href{https://doi.org/10.1051/0004-6361/201832666}{\emph{A\&A} {\bfseries
  616} (2018) A57} [\href{https://arxiv.org/abs/1801.06483}{{\ttfamily
  1801.06483}}].

\bibitem{Levinson2000}
A.~{Levinson}, \emph{{Particle Acceleration and Curvature TeV Emission by
  Rotating, Supermassive Black Holes}},
  \href{https://doi.org/10.1103/PhysRevLett.85.912}{\emph{Phys Rev Lett}
  {\bfseries 85} (2000) 912}.

\bibitem{Neronov2009}
A.~Y. {Neronov}, D.~V. {Semikoz} and I.~I. {Tkachev}, \emph{{Ultra-high energy
  cosmic ray production in the polar cap regions of black hole
  magnetospheres}},
  \href{https://doi.org/10.1088/1367-2630/11/6/065015}{\emph{New Journal of
  Physics} {\bfseries 11} (2009) 065015}
  [\href{https://arxiv.org/abs/0712.1737}{{\ttfamily 0712.1737}}].

\bibitem{Rieger2011}
F.~M. {Rieger}, \emph{{Nonthermal Processes in Black Hole-Jet Magnetospheres}},
  \href{https://doi.org/10.1142/S0218271811019712}{\emph{International Journal
  of Modern Physics D} {\bfseries 20} (2011) 1547}
  [\href{https://arxiv.org/abs/1107.2119}{{\ttfamily 1107.2119}}].

\bibitem{Ptitsyna2016}
K.~{Ptitsyna} and A.~{Neronov}, \emph{{Particle acceleration in the vacuum gaps
  in black hole magnetospheres}},
  \href{https://doi.org/10.1051/0004-6361/201527549}{\emph{A\&A} {\bfseries
  593} (2016) A8} [\href{https://arxiv.org/abs/1510.04023}{{\ttfamily
  1510.04023}}].

\bibitem{Moncada2017}
R.~J. {Moncada}, R.~A. {Colon}, J.~J. {Guerra}, M.~J. {O'Dowd} and L.~A.
  {Anchordoqui}, \emph{{Ultrahigh energy cosmic ray nuclei from remnants of
  dead quasars}},
  \href{https://doi.org/10.1016/j.jheap.2017.04.001}{\emph{Journal of High
  Energy Astrophysics} {\bfseries 13} (2017) 32}
  [\href{https://arxiv.org/abs/1702.00053}{{\ttfamily 1702.00053}}].

\bibitem{Rieger2018}
F.~{Rieger} and A.~{Levinson}, \emph{{Radio Galaxies at VHE Energies}},
  \href{https://doi.org/10.3390/galaxies6040116}{\emph{Galaxies} {\bfseries 6}
  (2018) 116} [\href{https://arxiv.org/abs/1810.05409}{{\ttfamily
  1810.05409}}].

\bibitem{Pedaletti2011}
G.~{Pedaletti}, S.~J. {Wagner} and F.~M. {Rieger}, \emph{{Very High Energy
  {\ensuremath{\gamma}}-ray Emission from Passive Supermassive Black Holes:
  Constraints for NGC 1399}},
  \href{https://doi.org/10.1088/0004-637X/738/2/142}{\emph{ApJ} {\bfseries 738}
  (2011) 142} [\href{https://arxiv.org/abs/1107.0910}{{\ttfamily 1107.0910}}].

\bibitem{Rieger2019}
F.~M. {Rieger}, \emph{{An Introduction to Particle Acceleration in Shearing
  Flows}}, {\emph{Galaxies} {\bfseries 7} (2019) 3}
  [\href{https://arxiv.org/abs/1909.07237}{{\ttfamily 1909.07237}}].

\bibitem{Ostrowski1998}
M.~{Ostrowski}, \emph{{Acceleration of ultra-high energy cosmic ray particles
  in relativistic jets in extragalactic radio sources}}, {\emph{A\&A}
  {\bfseries 335} (1998) 134}
  [\href{https://arxiv.org/abs/astro-ph/9803299}{{\ttfamily
  astro-ph/9803299}}].

\bibitem{Ostrowski2000}
M.~{Ostrowski}, \emph{{On possible `cosmic ray cocoons' of relativistic jets}},
  \href{https://doi.org/10.1046/j.1365-8711.2000.03146.x}{\emph{MNRAS}
  {\bfseries 312} (2000) 579}
  [\href{https://arxiv.org/abs/astro-ph/9910491}{{\ttfamily
  astro-ph/9910491}}].

\bibitem{Caprioli2015}
D.~{Caprioli}, \emph{{``Espresso'' Acceleration of Ultra-high-energy Cosmic
  Rays}}, \href{https://doi.org/10.1088/2041-8205/811/2/L38}{\emph{ApJL}
  {\bfseries 811} (2015) L38}
  [\href{https://arxiv.org/abs/1505.06739}{{\ttfamily 1505.06739}}].

\bibitem{Kimura2018}
S.~S. {Kimura}, K.~{Murase} and B.~T. {Zhang}, \emph{{Ultrahigh-energy
  cosmic-ray nuclei from black hole jets: Recycling galactic cosmic rays
  through shear acceleration}},
  \href{https://doi.org/10.1103/PhysRevD.97.023026}{\emph{Phys Rev D}
  {\bfseries 97} (2018) 023026}
  [\href{https://arxiv.org/abs/1705.05027}{{\ttfamily 1705.05027}}].

\bibitem{Rieger2004}
F.~M. {Rieger} and P.~{Duffy}, \emph{{Shear Acceleration in Relativistic
  Astrophysical Jets}}, \href{https://doi.org/10.1086/425167}{\emph{ApJ}
  {\bfseries 617} (2004) 155}
  [\href{https://arxiv.org/abs/astro-ph/0410269}{{\ttfamily
  astro-ph/0410269}}].

\bibitem{Rieger2016}
F.~M. {Rieger} and P.~{Duffy}, \emph{{Shear Acceleration in Expanding Flows}},
  \href{https://doi.org/10.3847/1538-4357/833/1/34}{\emph{ApJ} {\bfseries 833}
  (2016) 34} [\href{https://arxiv.org/abs/1611.04342}{{\ttfamily 1611.04342}}].

\bibitem{Liu2017}
R.-Y. {Liu}, F.~M. {Rieger} and F.~A. {Aharonian}, \emph{{Particle Acceleration
  in Mildly Relativistic Shearing Flows: The Interplay of Systematic and
  Stochastic Effects, and the Origin of the Extended High-energy Emission in
  AGN Jets}}, \href{https://doi.org/10.3847/1538-4357/aa7410}{\emph{ApJ}
  {\bfseries 842} (2017) 39}
  [\href{https://arxiv.org/abs/1706.01054}{{\ttfamily 1706.01054}}].

\bibitem{Webb2018}
G.~M. {Webb}, A.~F. {Barghouty}, Q.~{Hu} and J.~A. {le Roux}, \emph{{Particle
  Acceleration Due to Cosmic-ray Viscosity and Fluid Shear in Astrophysical
  Jets}}, \href{https://doi.org/10.3847/1538-4357/aaae6c}{\emph{ApJ} {\bfseries
  855} (2018) 31}.

\bibitem{Webb2019}
G.~M. {Webb}, S.~{Al-Nussirat}, P.~{Mostafavi}, A.~F. {Barghouty}, G.~{Li},
  J.~A. {le Roux} et~al., \emph{{Particle Acceleration by Cosmic Ray Viscosity
  in Radio-jet Shear Flows}},
  \href{https://doi.org/10.3847/1538-4357/ab2fca}{\emph{ApJ} {\bfseries 881}
  (2019) 123}.

\bibitem{Lemoine2010}
M.~{Lemoine} and G.~{Pelletier}, \emph{{On electromagnetic instabilities at
  ultra-relativistic shock waves}},
  \href{https://doi.org/10.1111/j.1365-2966.2009.15869.x}{\emph{MNRAS}
  {\bfseries 402} (2010) 321}
  [\href{https://arxiv.org/abs/0904.2657}{{\ttfamily 0904.2657}}].

\bibitem{Sironi2011}
L.~{Sironi} and A.~{Spitkovsky}, \emph{{Particle Acceleration in Relativistic
  Magnetized Collisionless Electron-Ion Shocks}},
  \href{https://doi.org/10.1088/0004-637X/726/2/75}{\emph{ApJ} {\bfseries 726}
  (2011) 75} [\href{https://arxiv.org/abs/1009.0024}{{\ttfamily 1009.0024}}].

\bibitem{Bell2018}
A.~R. {Bell}, A.~T. {Araudo}, J.~H. {Matthews} and K.~M. {Blundell},
  \emph{{Cosmic-ray acceleration by relativistic shocks: limits and
  estimates}}, \href{https://doi.org/10.1093/mnras/stx2485}{\emph{MNRAS}
  {\bfseries 473} (2018) 2364}
  [\href{https://arxiv.org/abs/1709.07793}{{\ttfamily 1709.07793}}].

\bibitem{Matthews2019}
J.~H. {Matthews}, A.~R. {Bell}, K.~M. {Blundell} and A.~T. {Araudo},
  \emph{{Ultrahigh energy cosmic rays from shocks in the lobes of powerful
  radio galaxies}}, \href{https://doi.org/10.1093/mnras/sty2936}{\emph{MNRAS}
  {\bfseries 482} (2019) 4303}
  [\href{https://arxiv.org/abs/1810.12350}{{\ttfamily 1810.12350}}].

\bibitem{Perucho2007}
M.~{Perucho} and J.~M. {Mart{\'\i}}, \emph{{A numerical simulation of the
  evolution and fate of a Fanaroff-Riley type I jet. The case of 3C 31}},
  \href{https://doi.org/10.1111/j.1365-2966.2007.12454.x}{\emph{MNRAS}
  {\bfseries 382} (2007) 526}
  [\href{https://arxiv.org/abs/0709.1784}{{\ttfamily 0709.1784}}].

\bibitem{Perucho2019}
M.~{Perucho}, J.-M. {Mart{\'\i}} and V.~{Quilis}, \emph{{Long-term FRII jet
  evolution: clues from three-dimensional simulations}},
  \href{https://doi.org/10.1093/mnras/sty2912}{\emph{MNRAS} {\bfseries 482}
  (2019) 3718} [\href{https://arxiv.org/abs/1810.10968}{{\ttfamily
  1810.10968}}].

\end{thebibliography}\endgroup


\end{document}